\documentstyle[12pt]{article}

\newcommand{\eq}{\begin{equation}}
\newcommand{\en}{\end{equation}}
\newcommand{\spz}{\hspace{0.7cm}}

\newcommand{\bN}{{\bf N}}
\newcommand{\bZ}{{\bf Z}}

\newcommand{\bC}{{\bf C}}

\newcommand{\D}{\Delta}

\newcommand{\ih}{\hat{\imath}}
\newcommand{\jh}{\hat{\jmath}}
\newcommand{\ibar}{\bar{\imath}}

\newcommand{\cA}{{\cal A}}
\newcommand{\cB}{{\cal B}}
\newcommand{\cC}{{\cal C}}
\newcommand{\cD}{{\cal D}}
\newcommand{\cE}{{\cal E}}
\newcommand{\cF}{{\cal F}}
\newcommand{\cH}{{\cal H}}

\newcommand{\NP}[1]{Nucl.\ Phys.\ {\bf #1}}
\newcommand{\PL}[1]{Phys.\ Lett.\ {\bf #1}}

\newcommand{\CMP}[1]{Comm.\ Math.\ Phys.\ {\bf #1}}

\newcommand{\IJMP}[1]{Int.\ J.\ Mod.\ Phys.\ {\bf #1}}

\begin{document}

\renewcommand{\thefootnote}{\fnsymbol{footnote}}

\newpage
\setcounter{page}{1}
\vskip 1cm
\begin{center}
{\large {\bf TOWARDS A CLASSIFICATION OF FUSION RULE ALGEBRAS IN RATIONAL
CONFORMAL FIELD THEORIES}}\\
\vspace{1.5cm}
{\large M.\ Caselle$^1$, G.\ Ponzano$^1$ and F.\ Ravanini$^2$}\\
\vspace{0.7cm}
{\em $^1$ Dept. of Theoretical Physics - University of Torino\\
          and I.N.F.N. - Sezione di Torino\\
          Via P.Giuria 1, I-10125 TORINO, Italy}\\
\vspace{0.4cm}
{\em $^2$ Service de Physique Th{\'e}orique, C.E.A. - Saclay
          \footnote{Laboratoire de la Direction des Sciences de la Mati{\`e}re
                    du Commissariat \`a l'Energie Atomique}\\
          Orme des Merisiers, F-91191 Gif-sur-Yvette, France\\
          and\\
          I.N.F.N. - Sezione di Bologna, Italy}
\end{center}
\vspace{.7cm}

\renewcommand{\thefootnote}{\arabic{footnote}}
\setcounter{footnote}{0}

\begin{abstract}
We review the main topics concerning Fusion Rule Algebras (FRA)
of Rational Conformal Field Theories. After an exposition of their
general properties, we examine known results on the complete
classification for low number of fields ($\leq 4$). We then turn our
attention to FRA's generated polynomially by one (real) fundamental field,
for which a classification is known. Attempting to generalize this result,
we describe some connections between FRA's and Graph Theory. The possibility
to get new results on the subject following this ``graph'' approach is
briefly discussed.
\end{abstract}

\vspace{1.7cm}

\begin{flushright}
Saclay preprint SPhT/91-174\\
Torino preprint DFTT 46/91\\
November 1991
\end{flushright}

\vskip 1cm
{\small Talk given by one of us (F.R.) at the {\it Research Conference on
Advanced Quantum Field Theory and Critical Phenomena}, Como (Italy), June
17-21, 1991}
\newpage

\section{Introduction}

Conformal Field Theories (CFT) in two dimensions~\cite{bpz,rcft} have
raised a lot of interest owing to their connections with critical
phenomena and string theories. In particular, the Rational Conformal
Field Theories (RCFT's),
endowed with a finite number of primary fields
with respect to some chiral algebra, and characterized by
rational values of central
charge and conformal weights, have been studied extensively.
The attempts to describe and classify
these theories with the tools of 2d conformal analysis and
modular invariance have shown the
relevant role played by Fusion Rule Algebras
(FRA)~\cite{verl,kawai}, which encode the
algebraic properties of primary fields. In some sense FRA's can be
considered as the skeletons of RCFT's, and lots of results have been
obtained in the reconstruction of RCFT's from the knowledge of their
FRA's~\cite{cr,mukhi,vafa,cp,pol,markov,fuchs}.

One of the open problems in the context of RCFT's is to
find their basic underlying principles and structures. So far we may
rely essentially on
two approaches to tackle this problem.
\begin{itemize}
\item In the first one, quantum groups are proposed as the underlying
      algebraic structure of RCFT's~\cite{qgroups}, and this approach is
      particularly successfull in the case of WZW models, although
      its extension to general RCFT's is still unclear. In particular this
      line of thougth, when applied toward any classification program,
      shows severe limitations due to the fact
      that the classification of quantum groups themselves is still an
      open problem.
\item The second approach exploits the relationships between
      RCFT's and three-dimensional Chern-Simons theories~\cite{c-s};
      in particular the
      Hilbert space associated to a constant time slice with charges in the
      three dimensional theories is related to the space of conformal blocks
      of corresponding RCFT's.
\end{itemize}
It is quite remarkable that both these approaches meet and fuse together at
the level of FRA's. Hence, any progress in trying to understand, organize
and classify FRA's  could be an important first step toward a deeper
comprehension not only of RCFT's, but also of Quantum Groups and Chern-Simons
theories.

In the following we will review some known results about FRA's and will
discuss some open issues.  Let us anticipate which will be our strategy:
FRA's in RCFT's must satisfy several stringent constraints
coming from different directions: diophantine constraints due to the
fact that the structure constants are positive integers, ``modular''
constraints on the matrix $S$ which diagonalizes the FRA, ``duality''
constraints on the allowed conformal weights and central charge of the
underlying RCFT. The main idea is then to enforce all these constraints
together by writing the FRA's in terms of their 1-dim irreducible
represantations; these encode, in a non trivial way,
all the relevant properties of the FRA's and have a natural
interpretation from the point of view of the quantum group approach,
some of them being the {\em quantum dimensions}~\cite{qgroups,dv} of the
irreps of the underlying quantum group.

During the last few years an impressive amount of new
results in this context has been presented, and to review all of
them would be not only beyond the scope of this contribution but also
beyond our forces. Therefore we hope to give at least a flavour
of the progress made in this field by reviewing the main
results obtained by various groups, trying to emphasize all the
connections among different approaches, and striving
to be, as far as possible, self contained.

This paper is organized as follows: sect.2 is devoted to a general
introduction on FRA's,  in sect.3 we discuss the classification of FRA's
for low number of primary fields, while sect.4 deals with the so called
{\em polynomial} FRA's introduced, studied and classified
in~\cite{pol} and later investigated (for
the particular case of WZW models) also in~\cite{gepner}. In sect.5 a
connection is traced between FRA's and Graph Theory. Finally sect.6 discusses
some open questions and other interesting results.

\section{Definitions and general setting}

A RCFT is a CFT whose physical Hilbert space decomposes into a
finite sum of highest weight irreducible representations (HWR) of the
(maximally extended~\cite{dv}) chiral algebra $\cA \otimes \overline{\cA}$
\eq
\cH = \bigotimes_{i,\ibar} \cH_i \otimes \overline{\cH}_{\ibar}
\en
where here and in the following (if not otherwise stated) the indices of the
middle of latin alphabet $i,j,k,...$ run on a {\em finite} set $X$ labelling
HWR's of $\cA$ at fixed central extension. This set can be put in one to one
correspondence with a set of integers ${0,1,2,...,r}$ such that $\cH_0$
denotes the representation whose highest weight state is the vacuum
 $|0\rangle$ of the
theory. One can consider the so called {\em chiral vertex
operators}~\cite{ms1,ms}, i.e. intertwiners among $\cH_i$'s
\eq
\Phi_{ij}^k(z)_t : \cH_i \otimes \cH_j \otimes \cH_k^* \to {\bf C}
\en
where $\cH_i^*$ is the dual space of $\cH_i$. Among $\cH_i$, $\cH_j$ and
$\cH_k^*$ there can be in general more that one possible coupling, i.e.
more that one
chiral vertex operator. The index $t$ distinguishes among them. Different
$\Phi_{ij}^k(z)_t$ for $t=1,...,N_{ij}^k$ span a vector space $V_{ij}^k$
of dimension $N_{ij}^k$. The numbers $N_{ij}^k$ are referred to as {\em
fusion rules}.

Consider formally an $r+1$ dimensional algebra $\cF$ over $\bC$ and a basis
$\Xi = \{\phi_0,\phi_1,...,\phi_r\}$ of it, such that in $\Xi$ the structure
constants are given exactly by the numbers $N_{ij}^k$
\eq
\phi_i \phi_j = \sum_k N_{ij}^k \phi_k
\en
The algebra $\cF$ is called {\em Fusion Rules Algebra} (FRA) associated with
the left chiral part of the given RCFT. The vectors $\phi_i$ are evidently in
1 to 1 correspondence with the HWR $\cH_i$, thus allowing to define a {\em
fusion product} between HWR's of $\cA$. The same procedure can be applied to
the right chiral algebra $\overline{\cA}$ to define fusion products between
HWR $\overline{\cH}_{\ibar}$ and a right fusion algebra $\overline{\cF}$

Properties of $\cF$ can be deduced from those of the chiral vertex operators.
These latter have been summarized in eqs.(4.17-4.18) of ref.~\cite{ms}; here
we simply mention those facts that justify the following assumptions for $\cF$:
\begin{itemize}
\item[{\bf P1}:]
     $\cF$ {\em is commutative}: $N_{ij}^k=N_{ji}^k$. This follows from the
     isomorphism (called $\Omega$ in~\cite{ms}) between $V_{ij}^k$ and
     $V_{ji}^k$.
\item[{\bf P2}:]
     $\cF$ {\em is associative}:
     \[
     \sum_m N_{ij}^m N_{km}^l = \sum_m N_{ik}^m N_{jm}^l
     \]
     This follows from existence of the {\em Fusion matrix} isomorphism
     \[
     F\left[\begin{array}{cc}i&k\\j&l\end{array}\right]~:~
     \bigoplus_m V_{im}^j \otimes V_{kl}^m \cong
     \bigoplus_m V_{ml}^j \otimes V_{ik}^m
     \]
     which is related to the assumption of duality.
\item[{\bf P3}:]
     {\em Identity}: there exists in $\Xi$ a vector $\phi_0$ called the
     identity such that for all $i,j\in X$: $N_{0i}^j=\delta_i^j$. The vector
     $\phi_0$ is the one in 1 to 1 correspondence with the HWR $\cH_0$.
\item[{\bf P4}:]
     {\em Charge conjugation}: there exists a one to one map of $X$, say
     $C:i\mapsto \ih$, such that $\hat{\ih}=i$ and $\cF$ is invariant under
     $C$, i.e. $N_{ij}^k=N_{\ih\jh}^{\hat{k}}$ and $N_{ij}^0=\delta_{i\jh}$.
     The matrix $C_{ij}\equiv N_{ij}^0$ uppers and lowers indices in
     $\cF$, and from {\bf P1,P2}, $N_{ijk}$ is totally symmetric in its
     indices.
\item[{\bf P5}:]
     {\em Integrality of structure constants}: $N_{ij}^k\in \bN$. This
     follows from the trivial fact that the dimension of a vector space
     like $V_{ij}^k$ is a non-negative integer.
\end{itemize}
In the following we shall adopt properties {\bf P1,...,P5} as postulates
defining (partially) the algebra $\cF$ and study the mathematical properties
of such an object.

The vectors $\phi_i$ are often called {\em fields}, to remember that they are
closely related to (chiral) primary fields of the underlying RCFT. A field
$\phi_{\ih}$ is referred to as the {\em conjugate} field to $\phi_i$. A FRA
such that $\ih=i$ for all $i\in X$ will be named {\em selfconjugate}.

A crucial role in the following is played by the {\em regular representation}
of $\cF$~\cite{kawai,graph} which associates to
$\phi_i$ the matrix $N_i$ of elements $(N_i)_j^k\equiv N_{ij}^k$.
It is a well known result in the mathematical theory of
associative, commutative algebras that all the $N_i$'s  can be
simultaneously diagonalized by a suitable  matrix $S$, i.e. all $N_i$'s
share the same set of $(r+1)$ non-null eigenvectors. In particular if
${\cal F}$ is selfconjugate, then all $N_i$'s are Hermitian, and $S$ is
orthogonal.
The $i$-th eigenvalue of $N_j$ will be denoted $\lambda_i(j)$.
The {\em spectrum} of ${\cal F}$, i.e. the set $\{\lambda_i(j)\}$,
encodes all the information about ${\cal F}$.
We recall the following properties:
\begin{itemize}
\item[{\bf T1}:]
     Being solutions of the algebraic equation (of degree $r+1$)
     \[
     det(N_j-\lambda_i(j){\bf 1})=0,
     \]
     with integer coefficients and with 1 as coefficient of the highest
     power, the $\lambda$'s are {\em algebraic integers}.
\item[{\bf T2}:]
     The $\lambda$'s provide all the irreps of ${\cal F}$; these latter
     being all one dimensional. In the $l$-th irrep, the role of
     $\phi_i$ is played by $\lambda_l(i)$, or
     \eq
     \lambda_l(i) \lambda_l(j)=
     \sum_{k=0}^rN_{ij}^k \lambda_l(k)
     \label{lambda}
     \en
\item[{\bf T3}:]
     As a consequence of (\ref{lambda}) $\lambda_l(\ih)= \lambda_l(i)^*$
     so that selfconjugate FRA's have real spectrum.
\item[{\bf T4}:]
     The spectrum satisfies the following {\em orthogonality
     relations}~\cite{kawai}:
     \eq
     \sum_{k=0}^r \lambda_i(k) \lambda_j(k)^*={1\over\nu_i}\delta_{ij}
     \label{orto1}
     \en
     \eq
     \sum_{k=0}^r  \nu_k\lambda_k(i) \lambda_k(j)^*=\delta_{ij}
     \label{orto2}
     \en
     where $\nu_i=(\sum_{k=0}^r|\lambda_i(k)|^2)^{-1}$. The meaning of
     the real numbers $\nu_i$'s will be clear later; notice that
     $0<\nu_i<1$ because $\lambda_j(0)=1$.
\item[{\bf T5}:]
     As a consequence of eqs.(\ref{lambda},\ref{orto2}) we can express
     the structure constants in terms of the $\lambda$'s as follows
     \[
     N_{ij}^k=\sum_{l=0}^r \nu_l\lambda_l(i) \lambda_l(j)
     \lambda_l(k)^*
     \]
     This clearly shows how $\lambda$'s really encode all the information
     about $\cF$.
\item[{\bf T6}:]
     The {\em unitary} matrix  which diagonalizes all $N_i$'s is
     \eq
     S_i^j=\sqrt{\nu_j}\xi_i^j
     \label{s}
     \en
     where $\nu_j$'s care for the normalization of the $j$-th
     eigenvector $\xi^j$: $N_k\xi^j=\lambda_j(k)\xi^j$, whose $l$-th
     component is related to $\lambda$'s by $\xi^j_l= (\xi^j)_l=
     \lambda_j(l)$.
\end{itemize}

Notice that different FRA's may actually describe the same algebraic
structure. We say that two FRA's ${\cal F}^{(1)},
{\cal F}^{(2)}$ are {\em isomorphic} if there is a {\em permutation}
$\pi: X \to X$ such that:
\begin{itemize}
\item $\phi_{\pi (i)}^{(1)}=\phi_{i}^{(2)}$ for all $i\in X$
\item $\pi(0)=\,0$
\item $[\pi,C]=0$
\item ${N^{(1)}}_{\pi(i)\pi(j)}^{\pi(k)}={N^{(2)}}_{ij}^{k}$ for all
      $i,j,k\in X$
\end{itemize}
Therefore FRA's of a given type are ``partitioned'' into {\em isomorphism
classes}, whose characterization is the final task of any classification
program.

A very interesting property of FRA's is encoded by the matrix $S$. It also
performs the modular transformation
$\tau\to-{1\over\tau}$ on the characters of the underlying
RCFT. This fact is known as {\em Verlinde theorem} in RCFT~\cite{verl,ms,dv}.
{}From our point of view, this amounts to a couple of new postulates to be
required on $\cF$:
\begin{itemize}
\item[{\bf P6}:]
     $S^2=C$. This, due to unitarity of $S$, is equivalent to the
     requirement that $S$ is symmetric, i.e. $\sqrt{\nu_j}\lambda_j(i)
     =\sqrt{\nu_i}\lambda_i(j)$, so that $\sqrt{\nu_j}=\sqrt{\nu_0}
     \lambda_0(j)$ and upon elimination of $\nu$'s:
     \eq
     \lambda_0(j)\lambda_j(i)=\lambda_0(i)\lambda_i(j)
     \label{s2}
     \en
     where $\lambda_0(j)\not= 0$, for all $r\in X$.
\item[{\bf P7}:]
     $(ST)^3=C$, where $T_{ij}=\delta_{ij}\exp 2\pi i (\Delta_i -
     c/24)$. where $\D_i$ is the conformal dimension attached to the
     HWR $\cH_i$ (i.e. the eigenvalue of the zero mode of the Virasoro
     subalgebra of $\cA$ for the highest weight vector of $\cH_i$) and $c$
     is the central charge of the same Virasoro subalgebra.
\end{itemize}

The first assumption {\bf P6} amounts to a very selective constraint on
possible FRA's. Many FRA's satisfying {\bf P1,...,P5} fail to satisfy this
constraint and must be discarded as candidates to build a RCFT.
The second {\bf P7} gives instead a set of equations to be
satisfied by the numbers $\{c,\D_i\}$. Some facts follow from these two new
assumptions:
\begin{itemize}
\item[{\bf T7}:]
     It is interesting to see that {\bf P6} and {\bf P7} can be combined in the
     following simple expressions~\cite{markov,qgroups}:
     \eq
     S_i^j=\sum_m S_0^m N^j_{im} e^{2i\pi (\D_i+\D_j-\D_m)}
     \label{simple}
     \en
     \eq
     e^{2\pi i\D_j}S_0^j=e^{\pi i c/4}\sum_m e^{2\pi i\D_m}S_0^mS_m^j
     \label{simple2}
     \en
\item[{\bf T8}:]
     {\em In any FRA of RCFT the eigenvalues $\lambda_i(j)$ are always
     linear combinations of roots of unity with integer coefficients}.
     Otherwise stated, $\lambda$'s not only are algebraic integers, but
     they belong to the algebraic integers of some ciclotomic field. This
     powerful and intriguing theorem has been proven in~\cite{markov}; the
     proof makes use of {\bf P1,...,P6}.
\end{itemize}

There is still an important issue to be discussed:
one has to require analytic closure of the
spaces of conformal blocks on Riemann surfaces of any genus and any number
of punctures. To find a complete formulation to this
requirement would be equivalent to solve (and classify) the whole set of
RCFT's. Here we shall be concerned with a more modest constraint that can be
deduced from a differential equation approach to analytic closure of the
space of conformal blocks of the 4-point function on the sphere~\cite{cr} and
of partition function on the torus~\cite{mukhi}. For an almost equivalent
constraint (although less powerful, as it does not give lower bounds on
$\D_i$'s and $c$) see~\cite{vafa}:
\begin{itemize}
\item[{\bf P8}:]
     Conformal dimensions $\D_i$ and central charge $c$ must satisfy the
     following set of equations:
     \begin{eqnarray*}
     &\displaystyle{\sum}_m&
     (N_{ij}^m N_{km}^l + N_{ik}^m N_{jm}^l + N_{il}^m N_{kj}^m)\D_m
     - N_{ijk}^l(\D_i+\D_j+\D_k+\D_l)\\ &=& \frac{N_{ijk}^l (N_{ijk}^l-1)}{2}
     - R_{ijk}^l
     \end{eqnarray*}
     where $N_{ijk}^l=\sum_m N_{ij}^m N_{km}^l$ and $R_{ijk}^l$ are
     non-negative integers, and
     \eq
     c=\frac{24}{n}\sum_{i=0}^{n-1} \D_i - 2(n-1) + 4l
     \en
     where $n$ is the number of {\em indipendent} characters in the partition
     function on the torus and $l=0,2,3,...$.
     These equations can be deduced~\cite{cr,mukhi} as Fuchs conditions
     on the differential equations to be satisfied by conformal blocks.
\end{itemize}
What is surprising is that the equations of {\bf P8} are in general
incompatible with those of {\bf P7} (in the form (\ref{simple},
\ref{simple2})). Only when
compatibility holds, the FRA is acceptable as a candidate for RCFT. The
experience of low $r$ results~\cite{cp,trieste} shows that this compatibility
between {\bf P7} and {\bf P8} is an extremely selective requirement (see
for examples how it works in the classification of all FRA's for $r=1$ in
ref.~\cite{trieste} or for $r=2$ in ref.~\cite{cp}).

Once all these constraints are imposed, the number of FRA's reduces sensibly
and one can hope, encouraged by results for low $r$ (see sect.3), to get a
classification of all possible structures obeying {\bf P1,...,P8}. Although
this problem of classification seems at present to be a formidable one, some
progress has been made. This will be the subject of next sections.

Moreover, the classification of all FRA's is only a first step in the
classification program of RCFT's. Once the FRA is given, one is faced with
the {\em reconstruction problem}, i.e. how to find, classify and solve the
RCFT's sharing the given FRA. Some steps in this direction have been done too
(see for example~\cite{cr,mukhi,kiritsis}). Here we recall only a well known
result on how to glue left and right FRA's to get a reasonable theory: it has
been proven~\cite{dv} that, if the chiral algebra $\cA$ is maximally extended,
{\em the right FRA must be isomorphic to the left one,
and right labelling of fields can differ from left ones only by an automorphism
of the FRA itself}.

Let us conclude this introductory section with some remarks on the concept of
{\em quantum dimension} which arises in the context of the quantum group
approach to RCFT's and has a natural interpretation in the language of
FRA; it was introduced by Dijkgraaf, E.Verlinde and H.Verlinde~\cite{dv} as a
(regularized) definition of dimension for the {\em unitary} HWR's $\cH_i$ of
$\cA$. The recipe is to divide
the character $\chi_i(\tau)$ of the irrep by the character of
the identity ($\cH_0$) irrep and then to take the limit $\tau \to 0$
(the rationale behind this definition is that if the Hilbert space were
finite dimensional, then $\chi_i(0)$ would exactly count the number of states):
\eq
d_i\equiv \lim_{\tau\to 0}\frac{\chi_i(\tau)}{\chi_0(\tau)}.
\en
We can compute $d_i$ using the modular transformation $S$:
\eq
d_i= \lim_{\tau\to 0}\frac{\sum_j S^j_i\chi_j(-1/\tau)}
{\sum_k S^k_0\chi_k(-1/\tau)},
\en
then, since in unitary theories $\D_i\geq0$ and $\D_i=0$ only for the identity
field, we get
\eq
d_i=\frac{S^i_0}{S^0_0}=\lambda_0(i)
\en
which are the eigenvalues of the corresponding FRA. Remarkably enough,
in the case of a WZW model built on a group G, this quantum dimension
exactly coincides with a suitable definition (using the so called
{\em Markov trace}) of dimension of the highest weight irreducible
representation of the quantum deformation of the same group G (see for
instance the last of ref.s~\cite{qgroups}).

\section{Classification of FRA's with low $r$}

Solving the Diophantine equations introduced in the previous section is
in general a difficult task. Similarly it is usually very difficult to
identify and  reconstruct all the allowed RCFT's given a consistent FRA.
Notwithstanding this, there are  some simple, but non trivial,
cases in which this can be done. In particular it is possible to
classify completely all the FRA's with one field plus the identity ($r=1$)
and with two fields plus the identity ($r=2$); but the complexity
grows exponentially as
the number of operators increases and seem to forbid much progress in
this direction. Some numerical results have been obtained for $r=3$.

\subsection{Exact classification for $r=1$}

In this case the general form of the algebras satisfying {\bf P1,...,P5} is
\[
\phi_1\phi_1=\phi_0+n\phi_1 \spz,\spz n\in\bN
\]
All these algebras also satisfy {\bf P6}. {\bf P7} (in the form of
eq.(\ref{simple})), gives some constraints on $c$ and $\D_1$ that are
compatible with those coming from {\bf P8} only for $n=0,1$. Hence the full
classification of FRA's with $r=1$~\cite{trieste} is given by the list of
tab.1
\begin{table}
\caption{{\protect\small The set of  FRA's, central charges and conformal
weights compatible  with RCFT's with one operator plus the
identity. $m=0,2,3,...$ and $l=0,1,2,...$.}}
\begin{center}
\begin{tabular}{|c|c|l|}
\hline
Algebra            & $c$& $\D$ \\
\hline
\rule[-3mm]{0mm}{8mm}
$\phi_1\phi_1=\phi_0\hskip 0.9cm$
&$1+4m$ & $\frac{1}{4}+l$ \\
\rule[-3mm]{0mm}{8mm}
&$7+4m$ & $\frac{3}{4}+l$ \\
\hline
\rule[-3mm]{0mm}{8mm}
$\phi_1\phi_1=\phi_0+\phi_1$
&$\frac{2}{5}+4m$ & $\frac{1}{5}+l$ \\
\rule[-3mm]{0mm}{8mm}
&$\frac{14}{5}+4m$ & $\frac{2}{5}+l$ \\
\rule[-3mm]{0mm}{8mm}
&$-\frac{22}{5}+4m$ & -$\frac{1}{5}+l$ \\
\rule[-3mm]{0mm}{8mm}
&$\frac{26}{5}+4m$ & $\frac{3}{5}+l$ \\
\hline
\end{tabular}
\end{center}
\end{table}
In tab.2 some corresponding RCFT's are
identified. Introducing a notation which will be clear in the
following we call them $\cA_1$ and $\cB_1$ algebras. It is interesting to
notice that all the allowed theories are {\em algebraic}~\cite{mathur}.
\begin{table}
\caption{{\protect\small Some RCFT's with one field plus the identity.}}
\begin{center}
\begin{tabular}{|c|c|c|l|}
\hline
Algebra            & $c$ & $\D$
                           & Model \\
\hline
\rule[-3mm]{0mm}{8mm}
$\phi_1\phi_1=\phi_0$&  $1$
&  $\frac{1}{4}$ & SU(2) k=1 WZW \\
\rule[-3mm]{0mm}{8mm}
&  $7$
&  $\frac{3}{4}$ & $E_7$ k=1 WZW \\
\hline
\rule[-3mm]{0mm}{8mm}
$\phi_1\phi_1=\phi_0+\phi_1$
 &  $\frac{14}{5}$  &  $\frac{2}{5}$
                         & $G_2$ k=1 WZW \\
\rule[-3mm]{0mm}{8mm}
 &  $\frac{26}{5}$  &  $\frac{3}{5}$
                         & $F_4$ k=1 WZW \\
\rule[-3mm]{0mm}{8mm}
 &  $-\frac{22}{5}$  &  $-\frac{1}{5}$
                         & (5,2) Virasoro \\
\hline
\end{tabular}
\end{center}
\end{table}

\subsection{Exact classification for $r=2$}

In this case there are only three
algebras which satisfy all the constraints~\cite{cp}.
They are listed in tab.3
\begin{table}
\caption{{\protect\small The set of  FRA's, central charges and conformal
weights compatible  with {\protect\em unitary} RCFT's with two fields plus the
identity. $m=0,2,3,...$ and $l,n=0,1,2,...$.}}
\begin{center}
\begin{tabular}{|c|c|c|c|}
\hline
Algebra            & $c$& $\D_1$
                            & $\D_2$ \\
\hline
\rule[-3mm]{0mm}{8mm}
$\phi_1\phi_1=\phi_2$
& $2+8m$ & $\frac{1}{3}+l$ &  $\frac{1}{3}+l$ \\
\rule[-3mm]{0mm}{8mm}
$\phi_2\phi_2=\phi_1$&&&\\
\rule[-3mm]{0mm}{8mm}
$\phi_1\phi_2=\phi_0$
& $6+8m$ & $\frac{2}{3}+l$ &  $\frac{2}{3}+l$ \\
\hline
\rule[-3mm]{0mm}{8mm}
$\phi_1\phi_1=\phi_0+\phi_2\hskip 0.9cm$
&$\frac{20}{7}+8m$ & $\frac{1}{7}+l$ &  $\frac{5}{7}+n$ \\
\rule[-3mm]{0mm}{8mm}
$\phi_2\phi_2=\phi_0+\phi_1+\phi_2$&&&\\
\rule[-3mm]{0mm}{8mm}
$\phi_1\phi_2=\phi_1+\phi_2\hskip 0.9cm$
&$\frac{36}{7}+8m$ & $\frac{6}{7}+l$ &  $\frac{2}{7}+n$ \\
\hline
\rule[-3mm]{0mm}{8mm}
$\phi_1\phi_1=\phi_0+\phi_2$&&&\\
\rule[-3mm]{0mm}{8mm}
$\phi_2\phi_2=\phi_0\hskip 0.9cm$
&$\frac{1}{2}+m$ & $\frac{1}{16}+\frac{l}{8}$ &  $\frac{1}{2}+n$ \\
\rule[-3mm]{0mm}{8mm}
$\phi_1\phi_2=\phi_1\hskip 0.9cm$&&&\\
\hline
\end{tabular}
\end{center}
\end{table}
Some identified RCFT's, with the corresponding conformal weights and charges,
are listed in tab.4. In this case we have two selfconjugate
algebras:
$\cA_2$and $\cB_2$ and one non-selfconjugate which is isomorphic to the
${\bf Z}_3$ group.
\begin{table}
\caption{{\protect\small Some identified models for $r=2$.}}
\begin{center}
\begin{tabular}{|c|c|c|c|l|}
\hline
Algebra            & $c$ & $\D_1$
                           & $\D_2$
                           & Model \\
\hline
\rule[-3mm]{0mm}{8mm}

$\phi_1\phi_1=\phi_2$
 & 2 & $\frac{1}{3}$
                       & $\frac{1}{3}$
                       & SU(3) $k=1$ WZW \\
\rule[-3mm]{0mm}{8mm}

$\phi_2\phi_2=\phi_1$ &&&&\\

\rule[-3mm]{0mm}{8mm}
$\phi_1\phi_2=\phi_0$
                   & 6 & $\frac{2}{3}$
                       & $\frac{2}{3}$
                       & $E_6$ $k=1$ WZW \\
\hline
\rule[-3mm]{0mm}{8mm}
$\phi_1\phi_1=\phi_0+\phi_2\hskip 0.9cm$
&&&&\\
\rule[-3mm]{0mm}{8mm}
$\phi_2\phi_2=\phi_0+\phi_1+\phi_2$&  $-\frac{68}{7}$
&  $-\frac{2}{7}$ & $-\frac{3}{7}$ & (7,2) Virasoro \\
\rule[-3mm]{0mm}{8mm}
$\phi_1\phi_2=\phi_1+\phi_2\hskip 0.9cm$
&&&&\\
\hline
\rule[-3mm]{0mm}{8mm}
$\phi_1\phi_1=\phi_0+\phi_2$
 &  $\frac{1}{2}$  &  $\frac{1}{16}$
                         & $\frac{1}{2}$
                         & Ising \\
\rule[-3mm]{0mm}{8mm}
$\phi_2\phi_2=\phi_0\hskip 0.9cm$
                   & $\frac{3}{2}$ & $\frac{3}{16}$
                         & $\frac{1}{2}$
                         & SU(2) $k=2$ WZW \\
\rule[-3mm]{0mm}{8mm}
$\phi_1\phi_2=\phi_1\hskip 0.9cm$
                   & $\frac{2n+1}{2}$ & $\frac{2n+1}{16}$
                           & $\frac{1}{2}$
                           & SO($2n+1$) $k=1$ WZW \\
\rule[-3mm]{0mm}{8mm}
                   &  $\frac{31}{2}$ & $\frac{15}{16}$
                         & $\frac{3}{2}$
                         & $E_8$ $k=2$ WZW \\
\hline
\end{tabular}
\end{center}
\end{table}
While the classification of the allowed conformal weights and charges is
complete, a proof of the completeness of the reconstruction of RCFT's is
still lacking. Remarkably enough, also in this case
all the allowed theories are $algebraic$~\cite{cp}.

\subsection{Numerical results for $r=3$}

A preliminary, numerical, analysis of FRA's with 3 fields plus the
identity has been done following the strategy of ref~\cite{cp}. Besides
the direct products of $r=1$ algebras we have found that only the
expected $\cA_3,\cB_3$ and $\bZ_4$ algebras satisfy all the constraints and
are good candidates for RCFT's. But it must be noticed that lots of
interesting structures appear if one only imposes associativity and
the modular constraints.

\section{Polynomial fusion rule algebras}

Remarkably enough there is another situation in which
the Diophantine system simplifies, namely the case of FRA's whose fields
are generated polynomially in terms of only one fundamental field (these
algebras will be called in the following {\em polynomial fusion rule
algebras}: PFRA). They are quite important in the context of RCFT's,
because they describe all models somehow related with the $SU(2)$
Kac-Moody algebra, hence, among the others, also the minimal models of
Virasoro and Supervirasoro algebras and all the $SU(2)\otimes SU(2)/SU(2)$
cosets.

Let us restrict our attention to those FRA's which are self-conjugate and
such that all fields $\phi_i$, $2\leq i \leq r$,
are generated polynomially in terms of a ``fundamental'' field
$\phi \equiv \phi_1$~\cite{kawai,pol}:
\eq
\phi_i = p_i(\phi)
\label{pol1}
\en
where $p_i$ is a suitable polynomial with real coefficients and
$p_0(\phi)=1$, $p_1(\phi)=\phi$.
Using the idempotent decomposition of the
algebra~\cite{kawai}, one can easily prove that
\eq
\lambda_j{(i)} = p_i(\xi_j)
\label{pol2}
\en
where $\xi_j \equiv \lambda_j{(1)}$ are {\em real} and {\em distinct},
else, by eq.(\ref{s}), $S$ would
have two equal columns and would be singular.
Inserting this equation into (\ref{orto2}) we get
\eq
\sum_{k=0}^r
\nu_k p_i(\xi_k)p_j(\xi_k) = \int_a^b dx \sum_{k=0}^r \nu_k \delta(x-\xi_k)
p_i(x) p_j(x) = \delta_{ij}
\en
for a finite interval $[a,b]$ such that  $\xi_k\in [a,b],0\leq k\leq r$.
Hence the polynomials ${p_i(x),~x\in{\bf R},~i=0,...,r}$,
build up an orthonormal set with respect to
the positive definite {\em atomic} measure
$\mu(x)=\sum_k \nu_k \delta(x-\xi_k)$.

Any set of orthogonal polynomials in one real variable
satisfies a 3-term recurrence relation, which
can be put in the form
\eq
xp_i(x) = a_{i+1} p_{i+1}(x) + b_i p_i(x) + c_{i-1} p_{i-1}(x)
\label{recurr0}
\en
where $b_0=c_{-1}=0$, $a_1=c_0=1$ and $a_{r+1}\not= 0$ is arbitrary.
Owing to (\ref{pol1},\ref{pol2}), this relation can be interpreted
in terms of fusion rules of the field
$\phi \equiv \phi_1$ with the other fields
\eq
\phi_1 \phi_i = \sum_{k=0}^r
N_{1i}^k \phi_k = \delta_{i<r}a_{i+1} \phi_{i+1} + b_i \phi_i +
c_{i-1} \phi_{i-1},~1\leq
i\leq r,
\label{recurr}
\en
where $\delta_{i<r}=1$ if $i<r$,  $\delta_{i<r}=0$ if $i\geq r$.
Hence the form of the matrix $N_1$ must be {\em tridiagonal}, symmetric
(i.e. $c_i=a_{i-1}$) as the fields are all self-conjugate, and all
$c_i$'s must not vanish:
\[
N_1 = \pmatrix{0&1& & & & & \cr
               1&b_1&c_1&&&&\cr
               &c_1&b_2&c_2&&&\cr
               &&c_2&b_3&.&&\cr
               &&&.&.&.&\cr
               &&&&.&.&c_{r-1}\cr
               &&&&&c_{r-1}&b_r}
\]
The ensuing class of FRA's enjoys rather peculiar properties, as we will
see. Once the set of $2r-1$ non-negative integer numbers
$c_i, 1\leq i \leq r-1$, and $b_i, 1\leq i \leq r$, is given,
then we can generate the whole set of orthogonal
polynomials, and hence all the $N_{ij}^k$, thanks to the fact that the
matrices
$N_i$, that constitute the regular representation of the FRA, are given
by $N_i = p_i(N_1)$. Combining this fact with the recurrence relation
(\ref{recurr}), one
can get the following set of equations, which automatically
encode all the
associativity conditions:
\eq
c_{i} N_{i+1,j}^k = c_{k-1} N_{i,j}^{k-1} + \delta_{k<r} c_{k} N_{i,j}^{k+1}
+ (b_k - b_i) N_{i,j}^k - c_{i-1} N_{i-1,j}^k
\label{pol3}
\en
where, as $i$ increases from 1 to $r-1$, all choices for
$j,k$ are considered such that:
$i+1\leq k \leq j\leq r$.
Once $N_p,~p\leq i$ are known,
this recurrence relation gives the not yet known
elements of $N_{i+1}$.

Moreover one can find an explicit expression for the
polynomials $p_i$~\cite{pol}.

With these results it is possible, at least in principle, to gain a
complete control over the spectrum of ${\cal F}$, and therefore to write the
constraint from {\bf P6}
as a set of algebraic Diophantine equations in $b$'s and $c$'s only.

In particular, in~\cite{pol} using the previous results
we obtained a complete classification of all the PFRA's
with structure constants less or equal to one.

Since all the $c_i$'s must be nonzero, they are fixed to one in this
case and we are left with a simpler problem with only $r$ degrees of
freedom. By imposing associativity (i.e. solving eq.(\ref{pol3}))
we find 5 infinite series
plus an isolated solution; they are completely characterized by the
values of the $b$'s ($b_0=0$):
\begin{itemize}
  \item[$\cA_r$:] $b_i=0$, $i=1 \cdots r$;
  \item[$\cB_r$:] $b_i=0$, $i=1 \cdots r-1$, $b_r=1$;
  \item[$\cB_r^{\prime}$:] $b_i=1$, $i=1 \cdots r$;
  \item[$\cC_r$:] If  $r\geq 2$, $b_i=1$, $i=1 \cdots r-1$, $b_r=0$;
  \item[$\cD_r$:] If  $r=2k\geq 4$, $b_{2i}=0,\,\, b_{2i-1}=1$,
   $i=1 \cdots k$;
  \item[$\cE_4$:] If $r=4$: $b_1=1,\,\,b_2=0,\,\,b_3=1,\,\,b_4=1$.
\end{itemize}

These solutions are not independent:
one can show that the two series $\cB_r$ and $\cB_r^{\prime}$
are isomorphic; but
this is the only isomorphism met in this classification.
For every one of these algebras one can compute the $S$ matrix which
diagonalizes the $N_i$'s and explicitely check that only $\cA_r$
and $\cB_r$ fulfil {\bf P6}. They also satisfy compatibility between {\bf P7}
and {\bf P8}, hence they are good candidates for building RCFT's.

A numerical check of (\ref{pol3}) has been done also allowing $N_{ij}^k>1$.
The results seem to confirm that $\cA_r$ and $\cB_r$ are the only consistent
series of PFRA's.

\section{FRA's and Graph Theory}

It is interesting to notice that
the $N_1$ matrix of the $\cA_r$ PFRA is the incidence
matrix of the Dynkin diagram of the $A_{r+1}$ Lie algebra.
and that the $\cB_r$ series can be obtained as a
$\bf Z_2$ folding of the $A_{2r+2}$ diagrams~\cite{pol}.

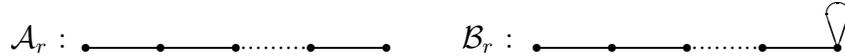
\setlength{\unitlength}{1mm}

\newsavebox{\tadpole}
\sbox{\tadpole}
{\begin{picture}(2,2)(0,0)
\put(0,0){\line(1,3){1.5}}
\put(0,0){\line(-1,3){1.5}}
\put(0,4.5){\oval(3,3)[t]}
\end{picture}}

\begin{center}
\begin{picture}(115,5)(0,0)
\put(0,0){$\cA_r$ :}
\multiput(10,0)(10,0){5}{\circle*{1}}
\multiput(10,0)(10,0){2}{\line(1,0){10}}
\multiput(31,0)(1,0){9}{\circle*{.2}}
\put(40,0){\line(1,0){10}}
\put(60,0){$\cB_r$ :}
\multiput(70,0)(10,0){5}{\circle*{1}}
\multiput(70,0)(10,0){2}{\line(1,0){10}}
\multiput(91,0)(1,0){9}{\circle*{.2}}
\put(100,0){\line(1,0){10}}
\put(110,0){\usebox{\tadpole}}
\end{picture}
\end{center}

This suggests a possible connection between PFRA's and Graph Theory.
This connection indeed exists and has been explored in a recent work by
De Boer and Goeree~\cite{markov}. Their results are in complete
agreement with ours. Moreover it is now clear that the impressive
simplifications which occur in PFRA's and the fact that it is possible
to obtain the above classification are
indeed signatures of the rich algebraic structure undelying
RCFT's~\cite{markov}.
Otherwise stated, the problem of classifcation of FRA's is related, in this
approach, to the problem of classifying matrices with non-negative integer
entries, a problem addressed in the book of Goodmann, De la Harpe and
Jones~\cite{ghj}. A classifcation of all non-negative integer valued matrices
whose highest component of the Perron-Frobenius eigenvector is less than
2 has been obtained and put in 1 to 1 correspondence with diagrams of type
$A_n$, $D_n$, $E_6$, $E_7$, $E_8$ and $A_{2n}/\bZ_2$. Now, if we check
{\bf P6} for all the FRA's generated by these diagrams, we discover that only
$A_n$ and $A_{2n}/\bZ_2$ survive the check. In other words, polynomiality of
FRA is equivalent to having highest component of Perron-Frobenius less than 2.
(See also the related topics in~\cite{frolich}).

Pushing the graph analogy further, one can ask if the FRA's of other known
series of models can be encoded in some graph. For example, the $D_{even}$
series of $SU(2)$ WZW models has FRA's with graphs:

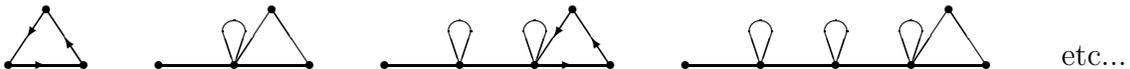
\newsavebox{\ftri}
\sbox{\ftri}
{\begin{picture}(10,10)(0,0)
\put(10,0){\circle*{1}}
\put(5,7.5){\circle*{1}}
\put(0,0){\vector(1,0){5}} \put(5,0){\line(1,0){5}}
\put(10,0){\vector(-2,3){2.5}} \put(7.5,3.75){\line(-2,3){2.5}}
\put(5,7.5){\vector(-2,-3){2.5}} \put(2.5,3.75){\line(-2,-3){2.5}}
\end{picture}}

\newsavebox{\tri}
\sbox{\tri}
{\begin{picture}(10,10)(0,0)
\put(10,0){\circle*{1}}
\put(5,7.5){\circle*{1}}
\put(0,0){\line(1,0){10}}
\put(10,0){\line(-2,3){5}}
\put(5,7.5){\line(-2,-3){5}}
\end{picture}}

\begin{center}
\begin{picture}(149,9)(0,0)

\put(0,0){\circle*{1}}
\put(0,0){\usebox{\ftri}}

\put(20,0){\circle*{1}}
\put(20,0){\line(1,0){10}}
\put(30,0){\circle*{1}}
\put(30,0){\usebox{\tadpole}}
\put(30,0){\usebox{\tri}}

\multiput(50,0)(10,0){3}{\circle*{1}}
\multiput(50,0)(10,0){2}{\line(1,0){10}}
\multiput(60,0)(10,0){2}{\usebox{\tadpole}}
\put(70,0){\usebox{\ftri}}

\multiput(90,0)(10,0){4}{\circle*{1}}
\multiput(90,0)(10,0){3}{\line(1,0){10}}
\multiput(100,0)(10,0){3}{\usebox{\tadpole}}
\put(120,0){\usebox{\tri}}

\put(140,0){etc...}
\end{picture}
\end{center}
These graphs can be obtained by {\em fusing} the $D_{2n}$ Dynkin
diagrams: a quite remarkable fact that relates them to the $A,D,E$
classification of~\cite{ghj}.

Other interesting classes of graphs can be obtained. For example $SU(3)$ WZW
models (diagonal series) have FRA's encoded in the series of graphs:
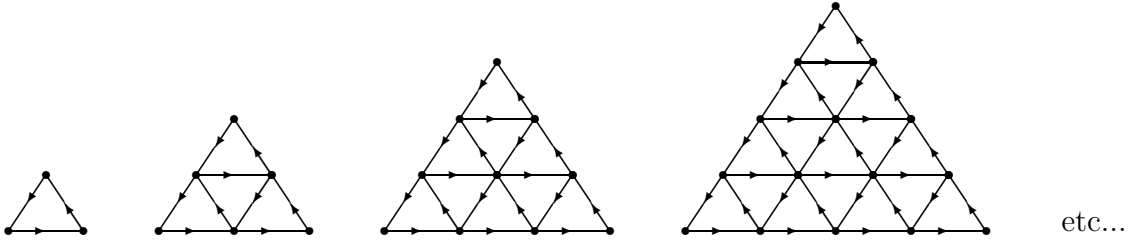
\begin{center}
\begin{picture}(149,30)(0,0)
\put(0,0){\circle*{1}}
\put(0,0){\usebox{\ftri}}

\put(20,0){\circle*{1}}
\multiput(20,0)(10,0){2}{\usebox{\ftri}}
\put(25,7.5){\usebox{\ftri}}

\put(50,0){\circle*{1}}
\multiput(50,0)(10,0){3}{\usebox{\ftri}}
\multiput(55,7.5)(10,0){2}{\usebox{\ftri}}
\put(60,15){\usebox{\ftri}}

\put(90,0){\circle*{1}}
\multiput(90,0)(10,0){4}{\usebox{\ftri}}
\multiput(95,7.5)(10,0){3}{\usebox{\ftri}}
\multiput(100,15)(10,0){2}{\usebox{\ftri}}
\put(105,22.5){\usebox{\ftri}}

\put(140,0){etc...}
\end{picture}
\end{center}
Finally, we give the graphs for the $G_2$ WZW models:
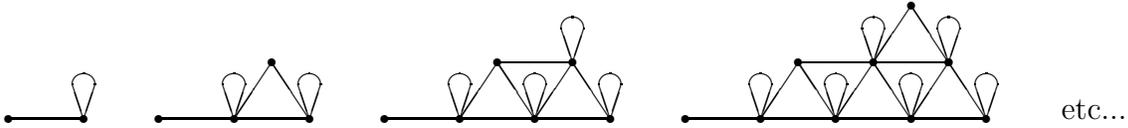
\begin{center}
\begin{picture}(149,15)(0,0)
\multiput(0,0)(10,0){2}{\circle*{1}}
\put(0,0){\line(1,0){10}}
\put(10,0){\usebox{\tadpole}}

\multiput(20,0)(10,0){2}{\circle*{1}}
\put(20,0){\line(1,0){10}}
\multiput(30,0)(10,0){2}{\usebox{\tadpole}}
\put(30,0){\usebox{\tri}}

\multiput(50,0)(10,0){2}{\circle*{1}}
\put(50,0){\line(1,0){10}}
\multiput(60,0)(10,0){3}{\usebox{\tadpole}}
\multiput(60,0)(10,0){2}{\usebox{\tri}}
\put(65,7.5){\line(1,0){10}}
\put(75,7.5){\usebox{\tadpole}}

\multiput(90,0)(10,0){2}{\circle*{1}}
\put(90,0){\line(1,0){10}}
\multiput(100,0)(10,0){4}{\usebox{\tadpole}}
\multiput(100,0)(10,0){3}{\usebox{\tri}}
\put(105,7.5){\line(1,0){10}}
\multiput(115,7.5)(10,0){2}{\usebox{\tadpole}}
\put(115,7.5){\usebox{\tri}}

\put(140,0){etc...}
\end{picture}
\end{center}
By suitably orbifolding, fusing and tensoring these graphs one should
obtain FRA's for
the non-diagonal series of WZW models, as well as those of minimal models of
the corresponding $W$-algebras. Work in this direction is in progress.

We conclude this section on applications of Graph Theory to FRA's by an
apparently unrelated but intriguing result on a systematic
analysis of FRA's with structure constants less or equal to 1,
imposing {\bf P1,...,P5} only, and without using constraints from modular
invariance or polynomiality (which are trivial to impose once this first
level classification has been obtained). The results are summarized in
tab.5.  The columns correspond to the number of fields (identity
included), while the row number $s\geq 0$ counts the pair of conjugate
fields. In fact as $C^2$ is the identity permutation on $\{0,1,
\cdots r\}$ the cycles of $C$ may have lenght 1,2 only, and non-selfconjugate
FRA's can be partitioned in a natural way according to the number $2s$
of non-selfconjugate fields.
\begin{table}
\caption{{\protect\small Enumeration of non-isomorphic algebras which satisfy
{\protect\bf P1,...,P5} and $N_{ij}^k\leq 1$. The entries show the number of
non-isomorphic associative algebras which are not direct products of lower
order ones; the subscripts specify the number (when different from 0) of
additional direct product algebras.}}
\begin{center}
\begin{tabular}{|c||c|c|c|c|c|c|c|}
\hline
    & \multicolumn{7}{c|}{$r+1$} \\
\cline{2-8}
$s$ & 1     & 2                & 3 & 4     & 5  & 6      & 7  \\
\hline \hline
0   & 1     & 2                & 3 & $3_3$ & 10 & $14_6$ & 18 \\
\hline
1   & \multicolumn{2}{c|}{ } & 1 & 4     & 5  & 9      & 14 \\
\hline
2   & \multicolumn{4}{c|}{ }             & 1  & $6_2$  & 7  \\
\hline
3   & \multicolumn{6}{c|}{ }                           & 1  \\
\hline
\end{tabular}
\end{center}
\end{table}

It is quite impressive to notice that, at least in the range
considered,  the number of FRA's as a
function of the number of operators grows very slowly, almost linearly.
In the case of selfconjugate FRA's, for instance, this has to be
compared with the asymptotic behaviour
\eq
H_r\sim \frac{2^{\scriptsize{\pmatrix{r+2\cr 3}}}}{r!}
\label{hr}
\en
of the number $H_r$ of non-isomorphic {\em hypergraphs}\footnote{They are
one-to-one with all subsets of the set formed by the unordered triples
$(iii),(iij),(ijk)$, $i\not=j\not=k$, $i\not=k$, $1\leq i,j,k\leq r$, and
in eq.~\ref{hr} the factor $(r!)^{-1}$ cares for the asymptotic
elimination of isomorphic algebras (see ref~\cite{harary}); for $r=5$
eq.~\ref{hr} is already accurate because it predicts $H_5\sim2.86\times
10^8$ while $H_5=287800704$ (from Polya's counting theorem).}
which underly these FRA's; $r=6$ yields $H_6\sim10^{14}$ and associativity
cuts it down to 18. The enumeration of associative hypergraphs seems to
be a non trivial problem, and it would be rather interesting to expand
tab.5 by resorting to supercomputing.

\section{Other results and open issues}

Most of the studies on FRA's have been done on the WZW models, in
particular those related to quantum groups~\cite{qgroups}. Moreover it
is widely believed that suitable cosets (and orbifolds) of WZW models
should exhaust all possible RCFT's. Finally the RCFT's which are more
important from the statistical mechanics point of view are definitely
the minimal models of the Virasoro Algebra and the $c=1$ models, and
all of them are well understood in terms of cosets (and orbifolds) of
Kac-Moody algebras. Lots of interesting exact results have been produced
in this direction.

\subsection{FRA's of WZW models}

All the FRA's for all possible (diagonal) WZW are known, at least in principle,
and can be explicitely written in some different but consistently
equivalent ways: using the so called {\em depth rule}~\cite{gep-wit}
(which was the first to be studied), using the Verlinde
formula~\cite{verl} and
the fact that modular transformation of WZW characters are well
known~\cite{kac};
more recently a highly efficient algorithm which mixes both these
approaches has been presented~\cite{fuchs-driel}.

\subsection{FRA's of Coset Models}

An interesting problem is that of finding the FRA's of the coset models
of the type:
\eq
\frac{G_k\times G_l}{G_{k+l}}
\label{coset}
\en
were $G$ can be  any Kac-Moody algebra and the index labels the level of
the algebra.
This point was discussed for $G=SU(2)$, $l=1$ in~\cite{pol} and for general
$G$'s and levels in~\cite{todorov}. If we
denote as ${\cal G}_k$ the FRA of the WZW
model (of the principal series)  with group $G$ and level $k$,
then the FRA of (\ref{coset}) proposed in~\cite{todorov} is
${\cal G}_k\bigotimes {\cal G}_l \bigotimes {\cal G}_{k+l}$.
However, to be more precise, if we want to write the actual FRA,
we have also to take into account
the foldings due to the so called {\em conformal grid symmetry}~\cite{bpz}.
This explaines why in~\cite{pol} we found in the case of Virasoro minimal
models~\cite{pol} the following rule: if $p$ and $q$ are
the coprimes which characterize the Virasoro model $(p,q)$ and $q$ is odd, then
the corresponding algebra is: $\cA_{p-2}\bigotimes \cB_{\frac{q-3}{2}}$.
Notice that, as a consequence of this, if both $p$ and $q$ are odd, we
have a non trivial
isomorphism between FRA's, i.e. $A_{p-2}\bigotimes B_{\frac{q-3}{2}}$
and $B_{p-2}\bigotimes A_{\frac{q-3}{2}}$ turn out to be isomorphic.
The same folding procedure should also be applied to FRA's of higher cosets
of the form (\ref{coset}).

\subsection{Relation with $N=2$ superconformal theory}

An interesting recent result has been obtained by
Gepner~\cite{gepner}, who has shown the deep relation existing between
FRA's and the operator product algebra of the chiral fields in $N=2$
superconformal field theories. A relation which is in principle
unexpected (notice that for $N=2$ we are dealing with the full operator
product algebra, without truncation of any kind), and could probably
lead to new results in the wider context of Frobenius Algebras and of
its potential deformations.

\vskip .3cm
All these issues show how the structure and classfication
of FRA's is a very intriguing problem
whose understanding will surely shed more light on the general properties
of CFT's and two dimensional Field Theories in general.

\vskip .5cm
\vskip 0.4 cm
{\bf Acknowledgements -} P.Christe and G.Mussardo are sincerely acknowledged
for the collaboration at the early stages of this project. We thank
J.Fr\"olich, J.Fuchs and J.-B.Zuber for useful discussions.
F.R. is grateful to the Theory Group of C.E.A.-Saclay for the kind hospitality
extended to him and to the Director of Sez. di Bologna of I.N.F.N., the
Theory Group of I.N.F.N.-Bologna and the Comm.IV of I.N.F.N. for the financial
support allowing him to spend this year in Saclay.

\end{document}